\documentclass[twocolumn,twoside]{revtex4}
\usepackage{graphicx}
\usepackage{fancyhdr}
\usepackage{calc}
\newcommand{\pythia}{\textsc{PYTHIA}}

\newcommand{\mathtx}{\mathrm}
\newcommand{\tev}{\mathtx{TeV}}
\newcommand{\gev}{\mathtx{GeV}}
\newcommand{\mev}{\mathtx{MeV}}

\newcommand{\pb}{\mathtx{pb}}
\newcommand{\nb}{\mathtx{nb}}
\newcommand{\ub}{\mathtx{\mu b}}

\newcommand{\Br}  {{\cal B}}
\newcommand{\sqrs}{\sqrt{s}}
\newcommand{\eold}{\sqrs =    7~\tev}
\newcommand{\enew}{\sqrs =    8~\tev}
\newcommand{\eall}{\sqrs =  7,8~\tev}
\newcommand{\elhc}{\sqrs =   13~\tev}

\newcommand{\absv}[1]{\left|#1\right|}
\newcommand{\absy}[1]{\absv{y_{#1}}}

\newcommand{\pt}[1]{p_{T,#1}}

\newcommand{\jpsi}{{J/\psi}}
\newcommand{\jps}[1]{{\psi(#1\mathtx{S})}}

\newcommand{\bz}{{B^0}}
\newcommand{\bd}{{B^0_d}}
\newcommand{\bs}{{B^0_s}}
\newcommand{\bp}{{B^+}}
\newcommand{\bu}{{B^\pm}}
\newcommand{\bq}{{B^0_{d,s}}}

\newcommand{\mm}{{\mu^+\mu^-}}
\newcommand{\pp}{{p \bar{p}}}
\newcommand{\kk}{{K^+K^-}}
\newcommand{\ku}{{K^\pm}}

\newcommand{\ks}{{K^{0}_S}}

\newcommand{\lz}{{\Lambda^0}}
\newcommand{\lb}{{\Lambda^0_b}}

\newcommand{\xsbn}{{\Xi_b^-}}

\newcommand{\bdsjpp}{{\bq \rightarrow \jpsi \pp}}
\newcommand{\bpjlbp}{{\bp \rightarrow \jpsi \bar{\Lambda}^0 p}}

\newcommand{\bujpku}{{\bu \rightarrow \jpsi \ku}}
\newcommand{\bsjphi}{{\bs \rightarrow \jpsi \phi}}

\newcommand{\brbjlp}{{\Br(\bpjlbp)}}

\newcommand{\npb}{Nucl. Phys. {\typesec B}}

\newcommand{\qrd}{Phys. Rev. {\typesec D}}
\newcommand{\qrl}{Phys. Rev. Lett.}
\newcommand{\plb}{Phys. Lett. {\typesec B}}
\newcommand{\epj}{Eur. Phys. J. {\typesec C}}

\newcommand{\hep}{J. High Energy Phys.}
\newcommand{\ins}{J. Inst.}
\newcommand{\cpc}{Comp. Phys. Comm.}

\newcommand{\cmscoll}{CMS Collaboration}

\newcommand{\atlcoll}{ATLAS Collaboration}
\newcommand{\lhbcoll}{LHCb Collaboration}

\newcommand{\eal}{\textit{et al.}}

\newcommand{\DOI}{}
\newcommand{\URL}{}
\newcommand{\aut}{}
\newcommand{\tit}{}
\newcommand{\jna}{}
\newcommand{\vol}{}
\newcommand{\pgn}{}
\newcommand{\yea}{}

\newcommand{\tpr}{\tit , }

\renewcommand{\tpr}{}

\newcommand{\dpr}[1]{\href{http://dx.doi.org/#1}{\typeurl{#1}}}

\newcommand{\typeaut}{}

\newcommand{\typetit}{}
\newcommand{\typeref}{}
\newcommand{\typesec}{}
\newcommand{\typevol}{\bf}
\newcommand{\typedoi}{}
\newcommand{\typeurl}{}

\begin{document}

\title{Production rates and branching fractions\\
of heavy hadrons \& quarkonia at LHC experiments}


\author{P.~Ronchese\\
on behalf of the ATLAS, CMS and LHCb collaborations}
\affiliation{University and INFN Padova}

\begin{abstract}
  Measurements of production cross-sections of inclusive $b$-hadrons pairs,
  bottom mesons and baryons, and quarkonia at LHC will be shown.
  Recent measurements of branching fractions of bottom baryons, bottom
  mesons in the final state with baryons, and a new result about a search
  for intermediate states in meson decay will also be shown.

\end{abstract}

\maketitle

\thispagestyle{fancy}


\section{Introduction}
Measurements of heavy hadron and quarkonia cross sections at LHC allow
probing QCD processes; they are also reference or ingredient for
searches and measurements of rarer or new processes, as well as the
baseline for associated production of heavy flavour and other objects.

The study of decay properties and branching fractions does allow
a test of form-factor models as well as the search for new and
exotic states that can be produced in the decay.

Results from
ATLAS~\cite{ref:aExper},
CMS~\cite{ref:cExper} and
LHCb~\cite{ref:lExper}
will be shown in the following.

\section{Production cross-sections}

\subsection{Inclusive $b$-hadrons}
An inclusive $b$-hadron pair production cross section measurement was
obtained by ATLAS~\cite{ref:aBBinc}
at an energy $\enew$; final states 
were selected looking for a $\jpsi$ coming from the first hadron and
decaying to $\mm$, and a muon coming from the second hadron.
The fiducial volume was defined requiring the two muons from the $\jpsi$
to have $\absv{\eta_{\mu,\jpsi}} < 2.3$ and the third muon to
have $\absv{\eta_\mu} < 2.5$; a minimum transverse momentum was also
required: $\pt{\mu} > 6 \gev$. The total cross section in the fiducial
volume was found:
%$\sigma(B(\rightarrow\jpsi[\rightarrow\mm]+X)B(\rightarrow\mu+X))) =
%(17.7 \pm 0.1(\mathtx{stat}) \pm 2.0(\mathtx{syst}))~\nb$
\begin{eqnarray*}
  \sigma(B(\rightarrow\jpsi[\rightarrow\mm]+X)B(\rightarrow\mu+X))) = & &\\
(17.7 \pm 0.1(\mathtx{stat}) \pm 2.0(\mathtx{syst}))~\nb\makebox[0pt][l]{~.}& &
\end{eqnarray*}

\subsection{Bottom mesons and baryons}

\subsubsection{$\bp$ production}

\begin{figure}[h]
\centering
\includegraphics[width=60mm]{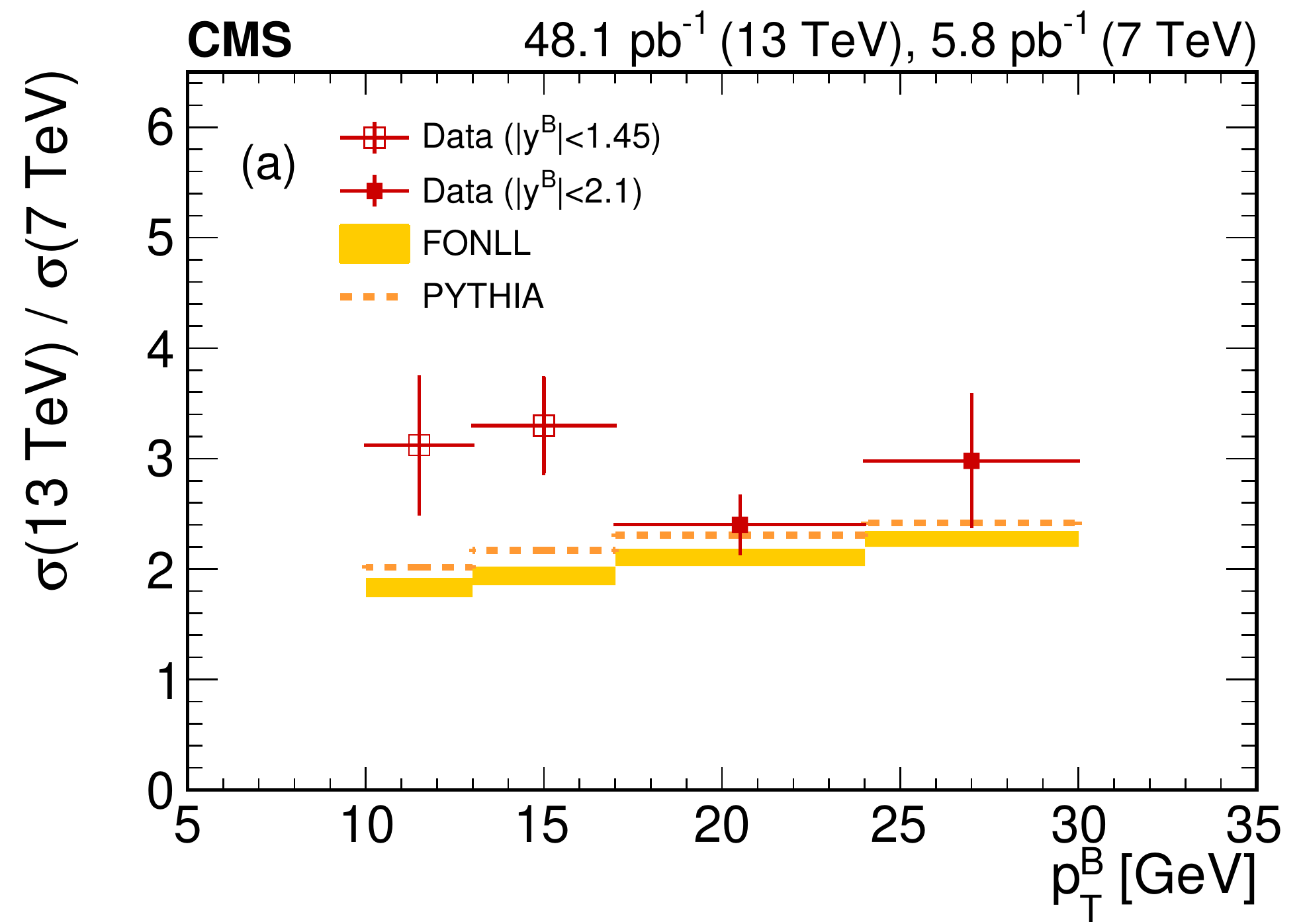}
\includegraphics[width=60mm]{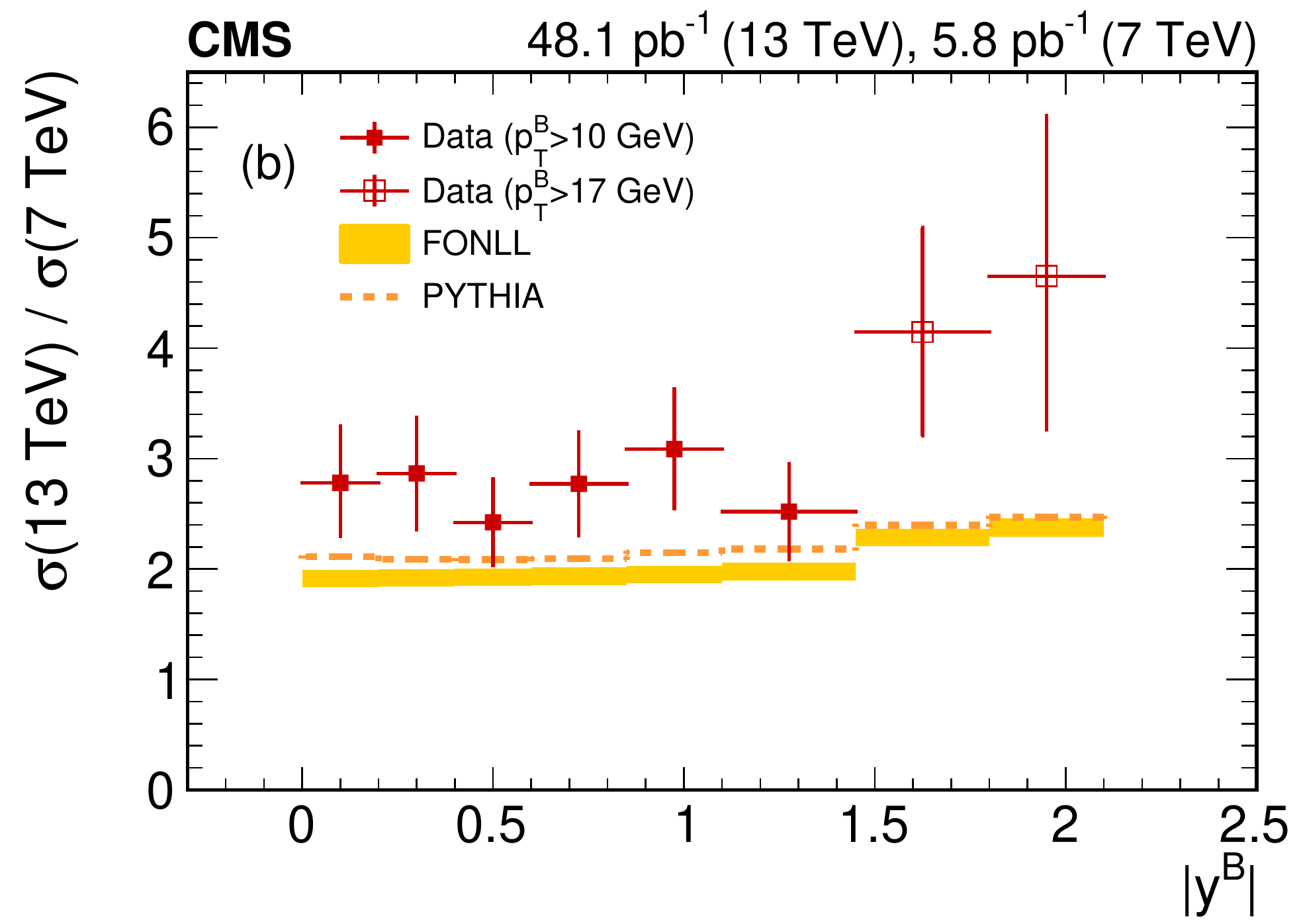}
\caption{Ratios of $\bp$ differential production cross sections at
  $\elhc$ and $\eold$ measured by CMS~\cite{ref:cBPdif}.}
\label{fig:cBPdif}
\end{figure}

The differential $pp \rightarrow \bp + X$ cross-section versus transverse
momentum or rapidity was measured by CMS~\cite{ref:cBPdif} at an
energy~$\elhc$ in the region $\absy{B} < 1.45$ or $\absy{B} < 2.1$ and
$10~\gev < \pt{B} < 100~\gev$ or $17~\gev < \pt{B} < 100~\gev$; the ratio
with the corresponding cross-section at~$\eold$ was measured 
and compared with FONLL~\cite{ref:fonll1,ref:fonll2,ref:fonll3}
and \pythia~\cite{ref:pythia} predictions.
Results are shown in Fig.~\ref{fig:cBPdif}. 

\begin{figure}[h]
\centering
\includegraphics[width=60mm]{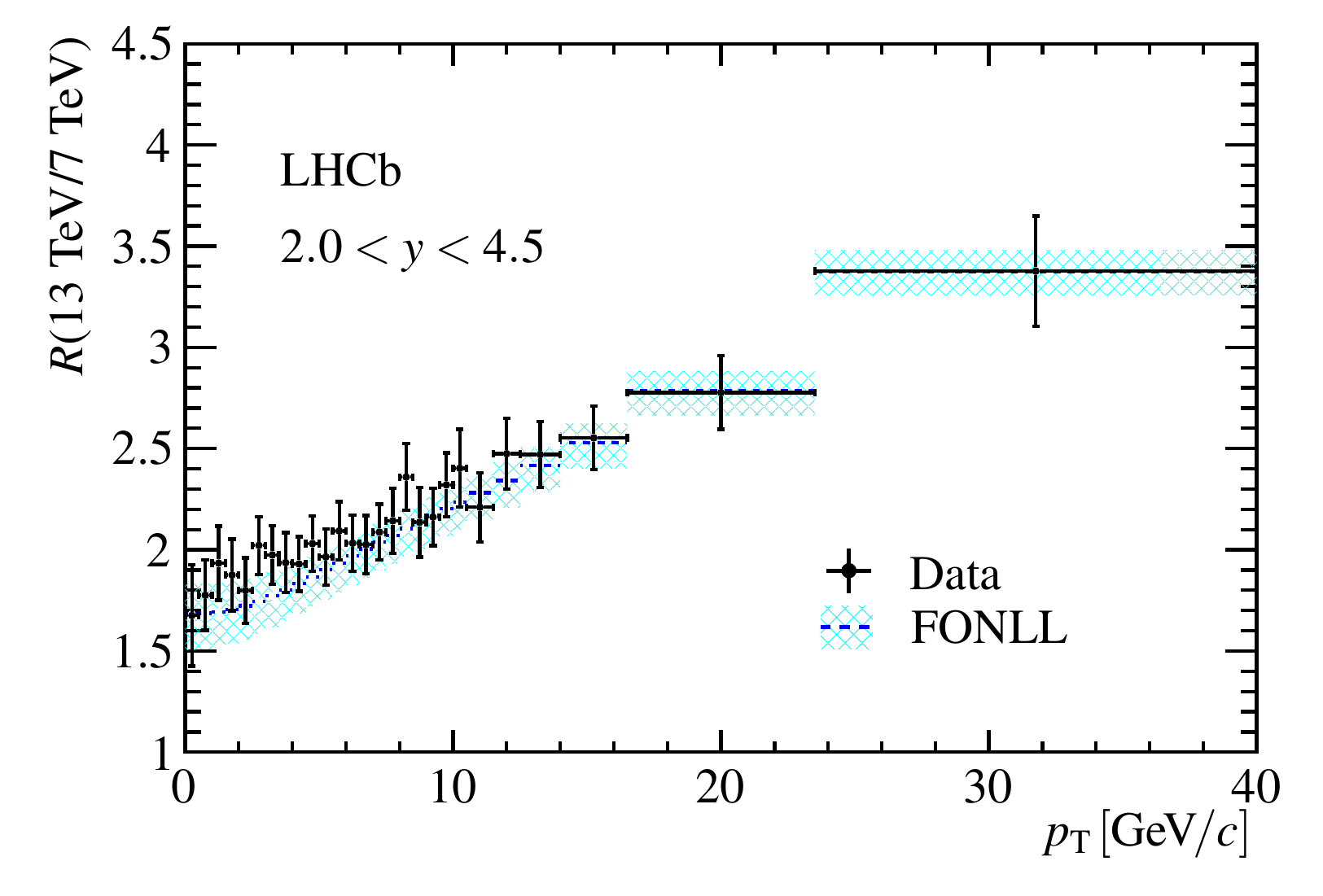}
\includegraphics[width=60mm]{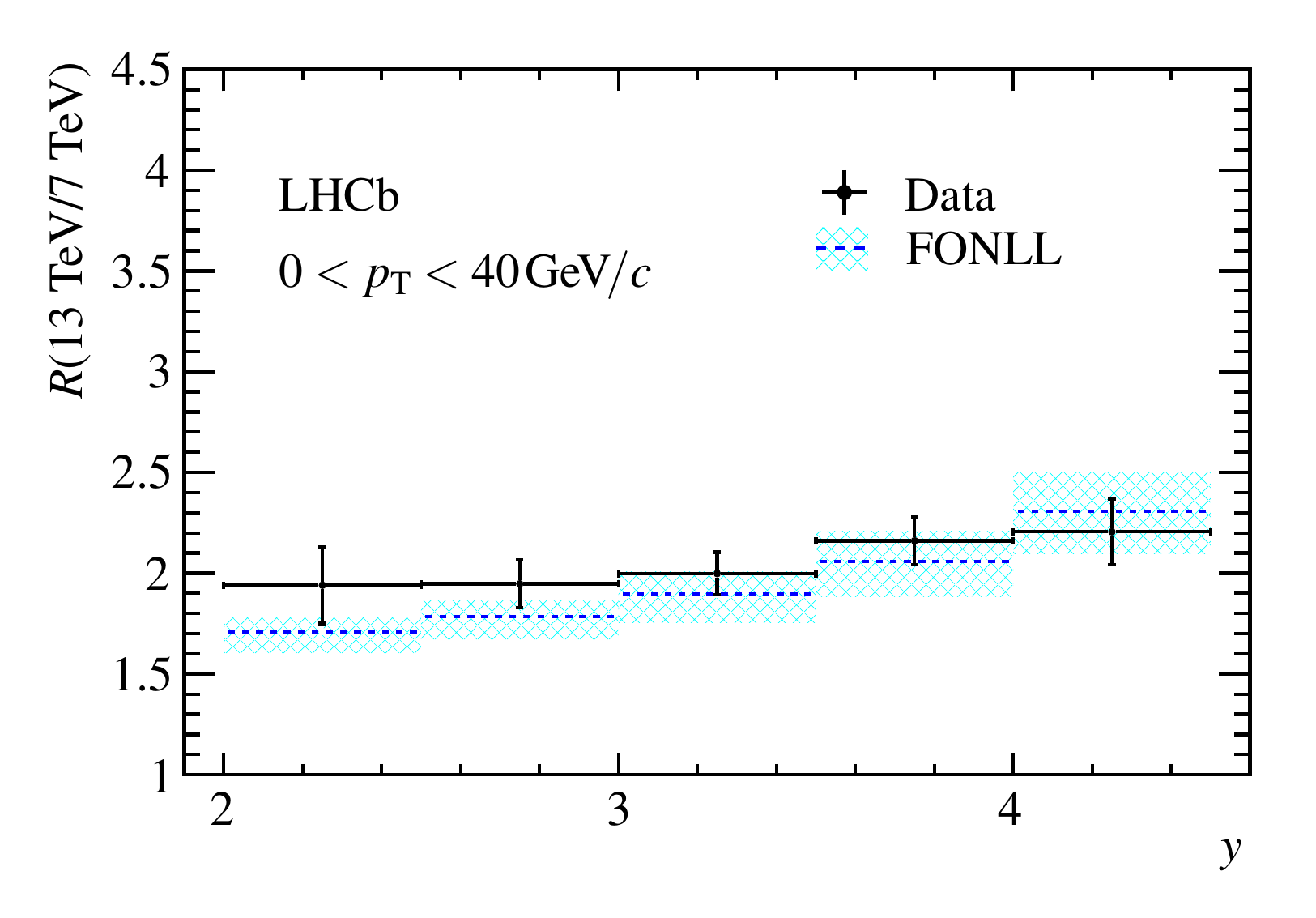}
\caption{Ratios of $\bp$ differential production cross sections at
  $\elhc$ and $\eold$ measured by LHCb~\cite{ref:lBPdif}.}
\label{fig:lBPdif}
\end{figure}

An analogous study was done by LHCb~\cite{ref:lBPdif}, that measured the
double differential cross-section versus transverse momentum and rapidity
in the region $2.0 < \absy{B} < 4.5$~, $\pt{B} < 40~\gev$. Again the ratio
with the corresponding cross-section at~$\eold$ was measured 
and compared with FONLL~\cite{ref:fonllf} predictions.
Differential cross-sections are shown in Fig.~\ref{fig:lBPdif}; the
integrated cross sections were found:
\begin{eqnarray*}
& &\sigma(pp \rightarrow \bu X~ (\sqrt{s} =~~7~\tev) =\\
& &( 43.0 \pm 0.2 (\mathtx{stat}) \pm 2.5 (\mathtx{syst}) \pm 1.7 (\mathtx{b.r.}) )~\ub\\
& &\sigma(pp \rightarrow \bu X~ (\sqrt{s} = 13~\tev) =\\
& &( 86.6 \pm 0.5 (\mathtx{stat}) \pm 5.4 (\mathtx{syst}) \pm 3.4 (\mathtx{b.r.}) )~\ub
\end{eqnarray*}
where the last uncertainty comes from the $\bujpku$ branching fraction.

\subsubsection{$\xsbn$ production}

Recently LHCb measured the ratio
$f_\xsbn / f_\lb$
%$\frac{f_\xsbn}{f_\lb}$
of the fragmentation
fractions to $\xsbn$ and $\lb$~\cite{ref:lFFRxl}. The decay of the
$\xsbn$~baryon has been studied since some time and a measurement of a
branching fraction was done~\cite{ref:lXbbrm}, but its absolute
determination requires knowing the fragmentation ratio.

\begin{figure}[h]
\centering
\includegraphics[width=60mm]{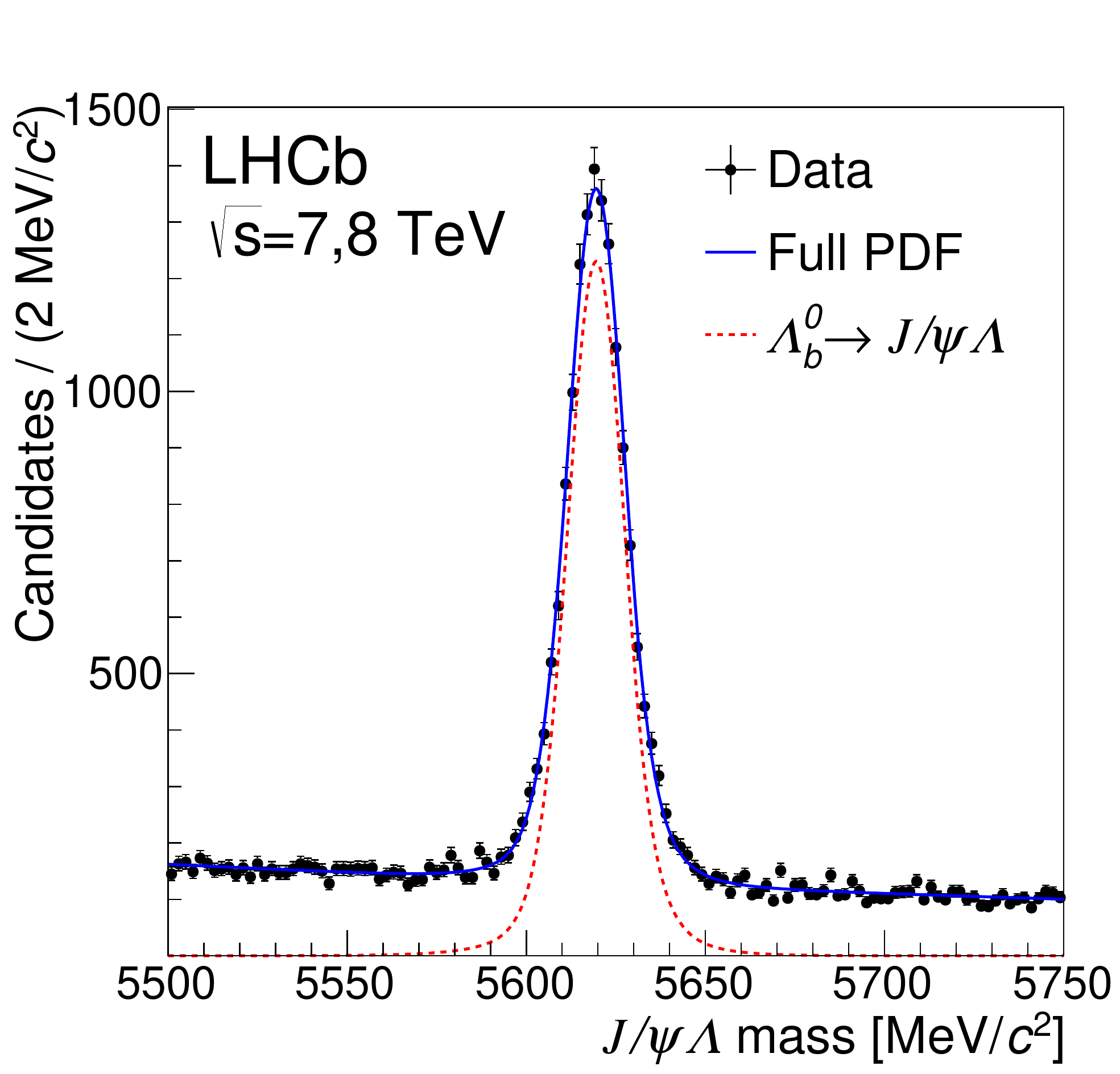}
\includegraphics[width=60mm]{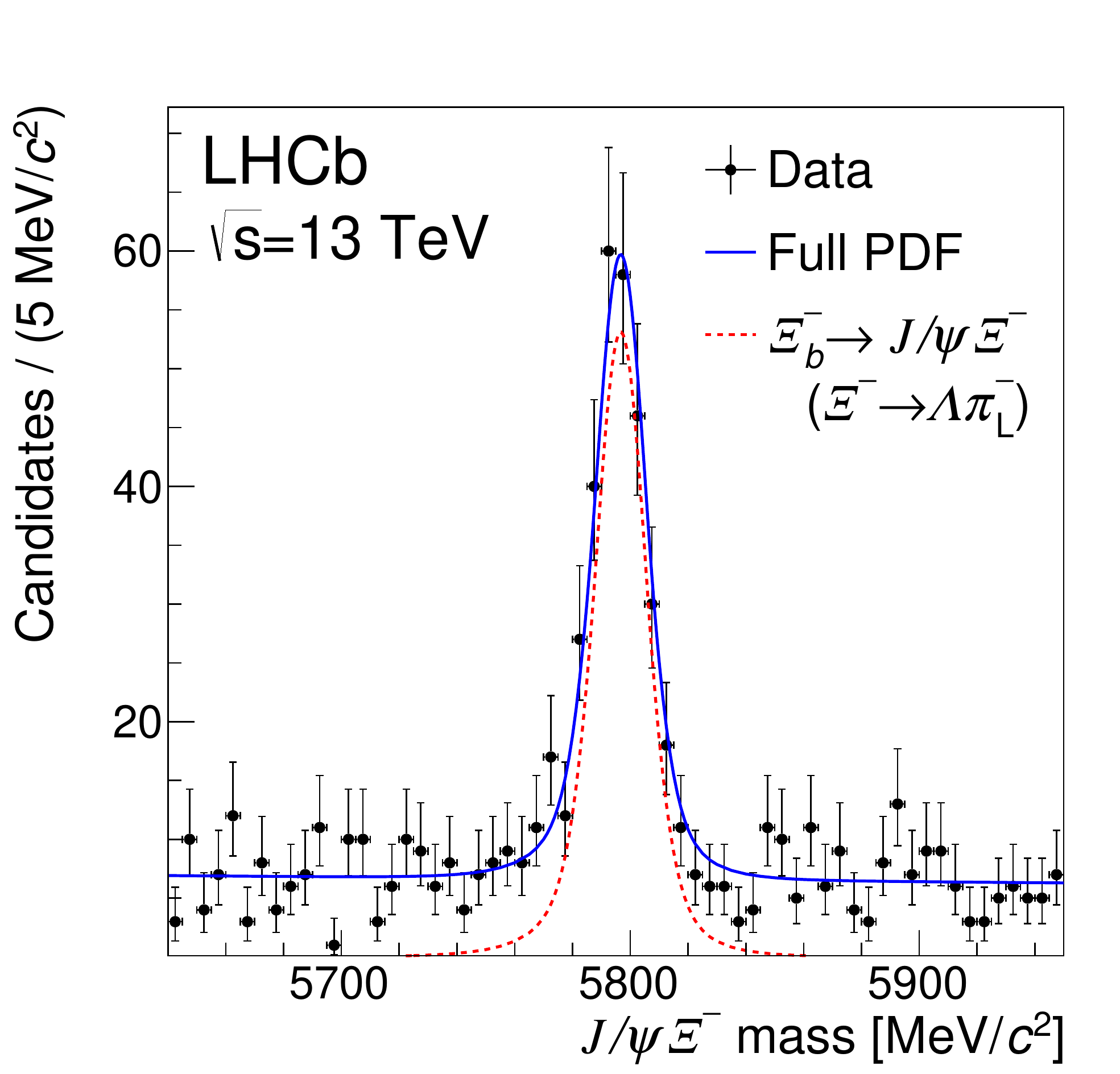}
\caption{Mass distributions for $\xsbn$ and $\lb$ reconstructed
  by LHCb~\cite{ref:lFFRxl}.}
\label{fig:lXbffr}
\end{figure}

The quantity that is directly accessible is the ratio $R$ of the
number of reconstructed decays in the
channel $\xsbn \rightarrow \jpsi \Xi^-$, and a
normalization one, with
$\lb   \rightarrow \jpsi \lz$,
but the ratio $R$ can be expressed also as the ratio of the products
of fragmentation and branching fractions; the latter can be expressed
as the products of the ratios of partial widths and lifetimes:
\begin{eqnarray*}
    R = \frac{f_\xsbn}{f_\lb} \cdot
        \frac{\Gamma(\xsbn \rightarrow \jpsi \Xi^-)}
             {\Gamma(\lb   \rightarrow \jpsi \lz)} \cdot
        \frac{\tau_\xsbn}
             {\tau_\lb}\makebox[0pt][l]{\rule{1pt}{0pt}} & &\\
      = \frac{N(\xsbn \rightarrow \jpsi \Xi^-)}
             {N(\lb   \rightarrow \jpsi \lz)} \cdot
        \frac{\epsilon_\lb}
             {\epsilon_\xsbn}\makebox[0pt][l]{~.} & &
\end{eqnarray*}
The ratio of widths can be assumed to be $2/3$ from $SU(3)$ flavor
symmetry~\cite{ref:su3fs1,ref:su3fs2,ref:su3fs3} 
and the ratio of lifetimes can be taken from PDG~\cite{ref:pdgPRD}
so that the fragmentation
ratio can be obtained.
The decay has been reconstructed pairing a $\jpsi \rightarrow \mm$ with a
$\lz \rightarrow p \pi^-$ or a $\Xi^- \rightarrow \lz \pi^-$. For the
reconstruction of the $\Xi^-$ or $\lz$ tracks have been classified as
``long'' or ``downstream'', depending on the track originating before
or after the vertex detector. For $\lz$ downstream tracks have
been used, due to the long lifetime, while a long track was used
as candidate for the pion coming from the $\Xi^-$ decay.
The signal yields were estimated from fits to the mass distributions,
as shown in Fig.~\ref{fig:lXbffr}.

Using the mass difference $m_\xsbn-m_\lb$ as free parameter in the fit
and using the $\lb$ mass from the PDG~\cite{ref:pdgPRD} the most
precise measurement of $\xsbn$ mass was obtained:
\begin{eqnarray*}
  & & m_\xsbn = \\
  & & (5796.70 \pm 0.39 (\mathtx{stat})
               \pm 0.15 (\mathtx{syst})
               \pm 0.17 (m_\lb))~\mev
\end{eqnarray*}
where the last uncertainty comes from the $\lb$ mass.

With the signal yields from the fits and the efficiencies from
simulation the ratio $R$ was obtained, and in the end the fragmentation
fraction was extracted:
\begin{eqnarray*}
  \frac{f_\xsbn}{f_\lb} = (6.7 \pm 0.5 (\mathtx{stat}) \pm 0.5 (\mathtx{syst}) \pm 2.0 (\mathtx{f.s.})) \times 10^{-2}& &\\
  {[}\makebox[\widthof{$\eall$}][l]{$\eall$}{]}& &\\
  \frac{f_\xsbn}{f_\lb} = (8.2 \pm 0.7 (\mathtx{stat}) \pm 0.6 (\mathtx{syst}) \pm 2.5 (\mathtx{f.s.})) \times 10^{-2}& &\\
  {[}\makebox[\widthof{$\eall$}][l]{$\sqrt{s} = \makebox[\widthof{$7,8~\tev$}][r]{$13~\tev$}$}{]}& &
\end{eqnarray*}
where the last uncertainty is due to the flavor symmetry assumption and
taken to be 30\%.

\subsection{Quarkonia}
Several measurements of the production cross-section for quarkonia have
been done at LHC experiments; a special interest can be found in the
production of quarkonia pairs. 
Quarkonia pairs can be produced in single parton scattering (SPS), that's
assumed to dominate and lead to strongly correlated pairs with small
rapidity differences, but, in the high parton densities in proton-proton
collisions, also double parton scattering (DPS) can occur producing
multiple heavy flavour particles with
large $\Delta y$~\cite{ref:qqpBar,ref:qqpKom}.
%The production
%mechanism can be tested looking at the transverse momentum distribution,
%with a color singlet component dominant at low $p_T$ and a color octet
%component dominant at high $p_T$.

\begin{figure}[h]
\centering
\includegraphics[width=60mm]{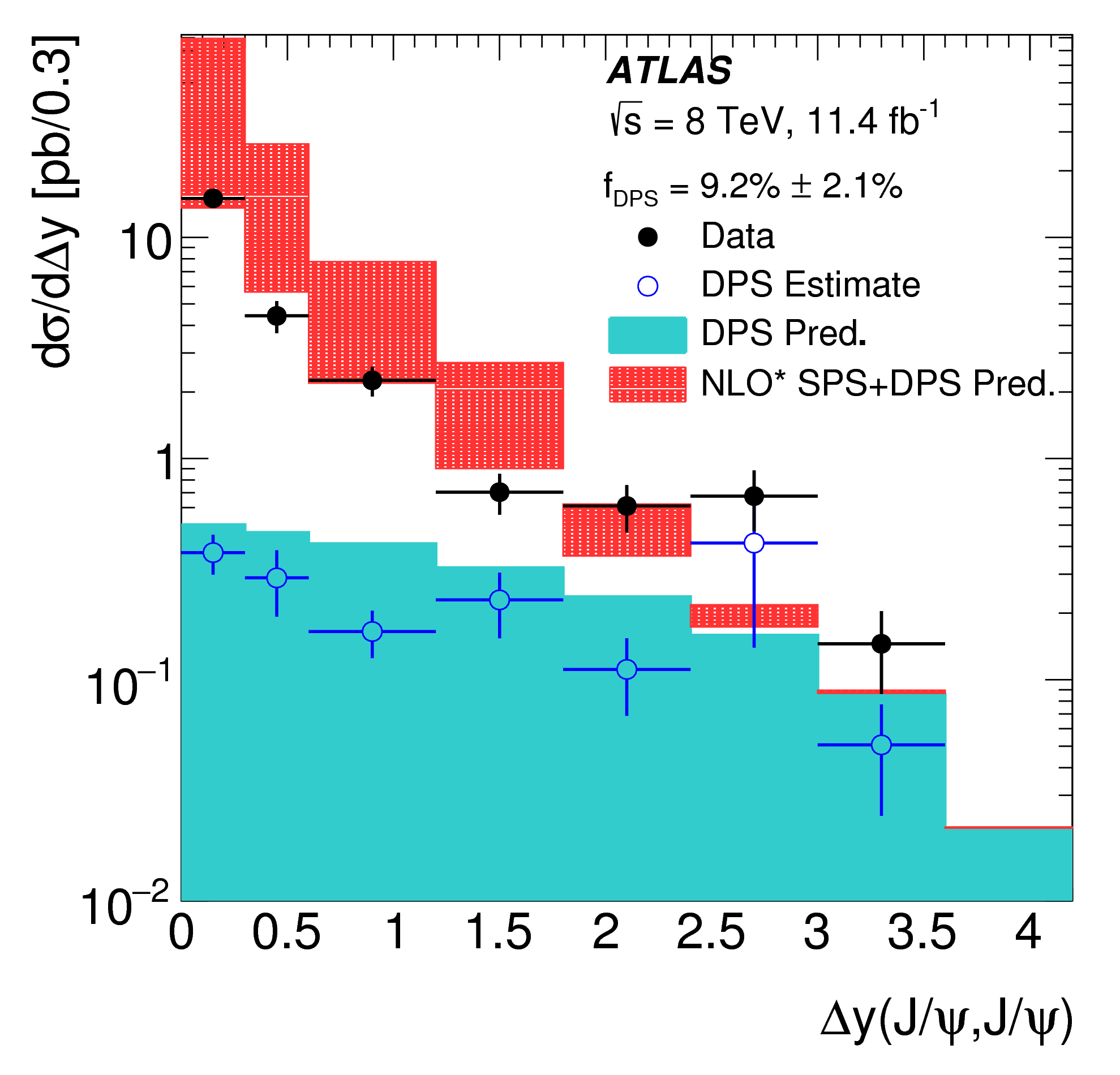}
\includegraphics[width=60mm]{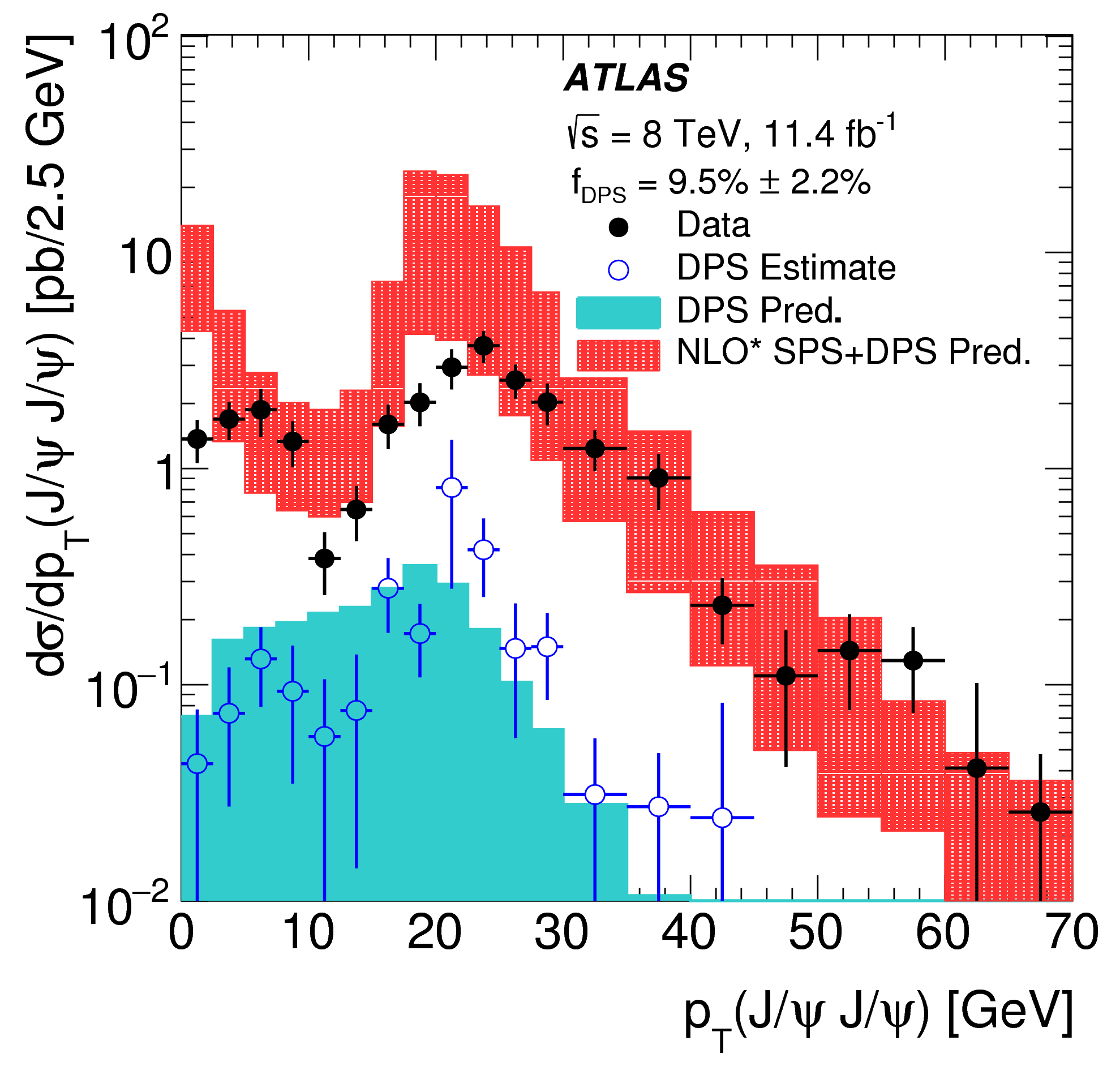}
\caption{The DPS and total differential cross-sections as a function of
  difference in rapidity and the transverse momentum of the di-$\jpsi$
  measured by ATLAS~\cite{ref:aDJPsi}, compared with
  LO DPS~\cite{ref:djDPSl} and
  NLO* SPS~\cite{ref:djSPS1,ref:djSPS2}
  predictions.}
\label{fig:aDJPsi}
\end{figure}

A measurement of the DPS contribution in double $\jpsi$ production was 
done by ATLAS~\cite{ref:aDJPsi} at~$\enew$. In the analysis $\jpsi$
pairs coming from different $pp$ interactions were removed with
a cut on the distance along the beam direction between the reconstructed
vertices, while the residual pile-up contamination was estimated looking
at the kinematic variables distributions in sidebands.
In double parton scattering $\jpsi$ candidates are assumed to be produced
independently, so a template $\Delta y \Delta \phi$ distribution has been 
built with $\jpsi$ pairs from different events and has been normalized
to data at large rapidity difference.
Then event weights in each $\Delta y \Delta \phi$ bin have been computed
from the ratio of the normalized template and full data; this weight can
be used as an estimate of the DPS fraction, that can be compared
to prediction from NLO versus rapidity difference and transverse momentum
as shown in Fig.~\ref{fig:aDJPsi}.

The total cross-section, in two fiducial regions $\absy{\jpsi} < 1.05$~,
$1.05 < \absy{\jpsi} < 2.1$ with $\pt{\jpsi} > 8.5~\gev$,
$\pt{\mu} > 2.5~\gev$ and $\absv{\eta_\mu} < 2.3$, were found to be:
%\newpage

\begin{eqnarray*}
  & &\sigma_\mathtx{Fid}(\makebox[\widthof{$1.05 < \absy{\jpsi} < \makebox[\widthof{$1.05$}][l]{$2.1$}$}][r]{$\absy{\jpsi} < \makebox[\widthof{$1.05$}][l]{$1.05$}$}) =\\
  & & (15.6 \pm 1.3 (\mathtx{stat}) \pm 1.2 (\mathtx{syst}) \pm 0.2 (\mathtx{b.r.}) \pm 0.3 (\mathtx{lum}))~\pb\\
  & &\sigma_\mathtx{Fid}(\makebox[\widthof{$1.05 < \absy{\jpsi} < \makebox[\widthof{$1.05$}][l]{$2.1$}$}][r]{$1.05 < \absy{\jpsi} < \makebox[\widthof{$1.05$}][l]{$2.1$}$}) =\\
  & & (13.5 \pm 1.3 (\mathtx{stat}) \pm 1.1 (\mathtx{syst}) \pm 0.2 (\mathtx{b.r.}) \pm 0.3 (\mathtx{lum}))~\pb
%      \makebox[267pt][r]{$\sigma_\mathtx{Fid} = (15.6 \pm 1.3 (\mathtx{st}) \pm 1.2 (\mathtx{sy}) \pm 0.2 (\mathtx{br}) \pm 0.3 (\mathtx{lum})) \pb$} & \makebox[58pt][r]{\tiny $\absy{\jpsi} < \makebox[\widthof{$1.05$}][l]{$1.05$}$}\\
%      \makebox[267pt][r]{$\sigma_\mathtx{Fid} = (13.5 \pm 1.3 (\mathtx{st}) \pm 1.1 (\mathtx{sy}) \pm 0.2 (\mathtx{br}) \pm 0.3 (\mathtx{lum})) \pb$} & \makebox[58pt][r]{\tiny $1.05 < \absy{\jpsi} < \makebox[\widthof{$1.05$}][l]{$2.1$}$}
\end{eqnarray*}

with a DPS fraction

\begin{eqnarray*}
  & &f_\mathtx{DPS} = (9.2 \pm 2.1 (\mathtx{stat}) \pm 0.5 (\mathtx{syst}))\%\makebox[0pt][l]{~.}
\end{eqnarray*}

%A similar measurement at $\elhc$ has been done from
%LHCb~\cite{ref:lDJPsi,ref:lDJPse}, that obtained DPS
%predictions from a large number of pseudoexperiments, where two uncorrelated
%$\jpsi$ mesons were produced according to the measured differential
%cross-sections, and SPS predictions from theoretical calculations
%using several approaches (LO, NLO, color singlet or color octet).
%Several data distributions were fitted with a two-component model to
%obtain the DPS fraction giving results in the range: 

A similar measurement at $\elhc$ has been done from
LHCb~\cite{ref:lDJPsi,ref:lDJPse}; the measured total cross-section
for the production in the region $2.0 < \absv{y} < 4.5$~, $p_T < 10~\gev$
was:
\begin{eqnarray*}
  \sigma = (15.2 \pm 1.0 (\mathtx{stat}) \pm 0.9 (\mathtx{syst}))~\nb\makebox[0pt][l]{~.}
\end{eqnarray*}

The DPS component prediction was obtained from a large number of
pseudoexperiments, where two uncorrelated
$\jpsi$ mesons were produced according to the measured differential
cross-sections, and SPS predictions from theoretical calculations
using several approaches (LO, NLO, color singlet or color octet).
Several data distributions were fitted with a two-component model to
obtain the DPS fraction giving results in the range: 

\begin{eqnarray*}
  f_\mathtx{DPS} = ((42 \pm 25) \div (86 \pm 55))\%\makebox[0pt][l]{~.}
\end{eqnarray*}

\begin{figure}[h]
\centering
\includegraphics[width=60mm]{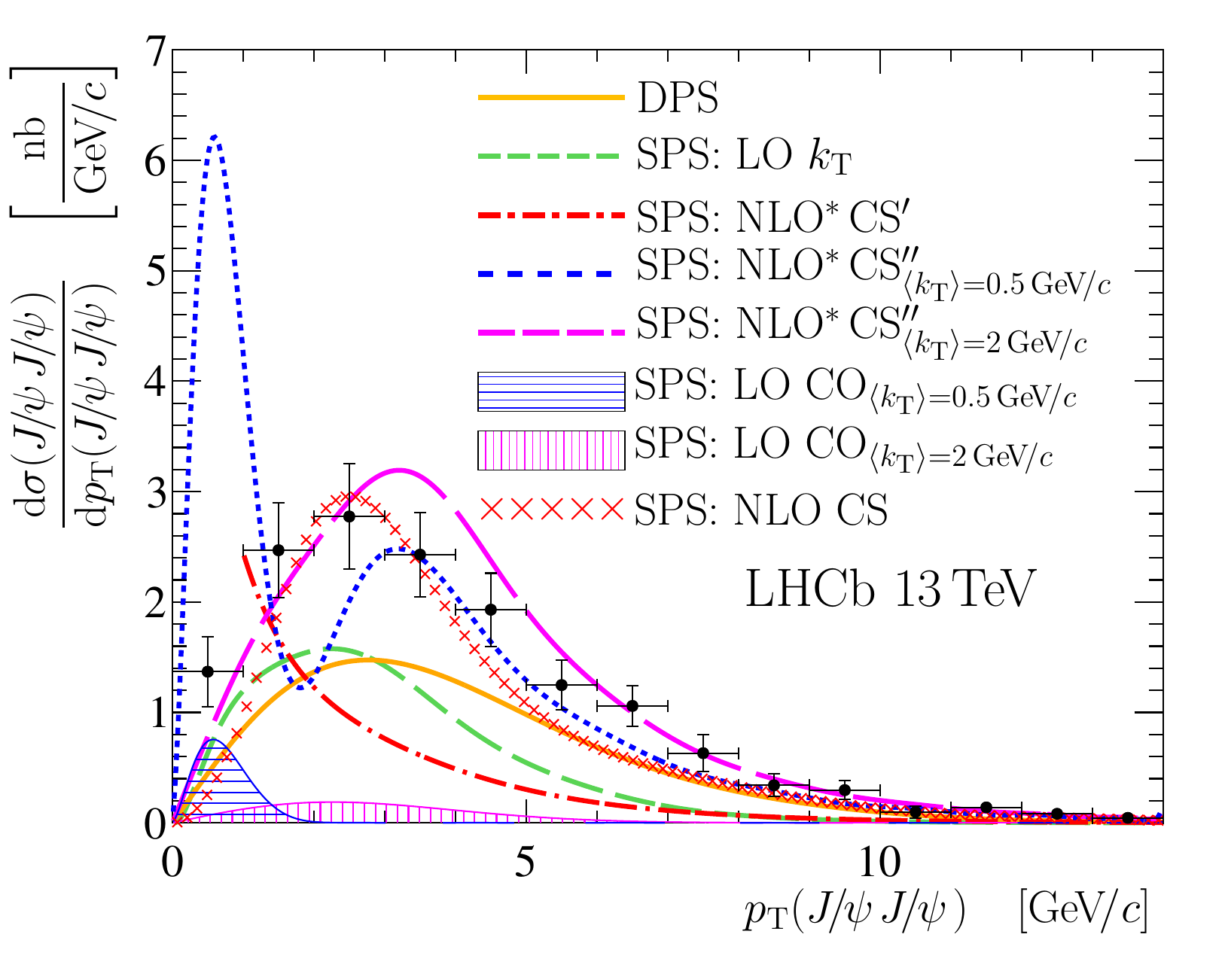}
\caption{Comparisons between the measured and predicted 
  differential cross-section vs. $p_{T,\jpsi\jpsi}$ from
  LHCb~\cite{ref:lDJPsi,ref:lDJPse}.}
\label{fig:lDJPsi}
\end{figure}

As an example, the comparison between the measured and predicted 
differential cross-section vs. $p_{T,\jpsi\jpsi}$ is shown in
Fig.~\ref{fig:lDJPsi}.

\section{Branching fractions}

\subsection{Bottom baryon decay}
Ratios of branching fractions of $b$-hadrons with a $\jpsi$ or a $\jps{2}$ 
in the final state allow testing the factorization of amplitudes; some recent
result of such ratios involve baryons.
A measurement of the ratio of branching fractions of
$\lb \rightarrow \jps{2} \lz$ and
$\lb \rightarrow \jpsi \lz$
was done by ATLAS~\cite{ref:aBRLbr}
a few years ago giving a result that shows a discrepancy from covariant
quark model prediction~\cite{ref:brLBr1,ref:brLBr2}.
Another measurement
has just been done by LHCb~\cite{ref:lBRLbr}
that reconstructed $\psi$ from non-prompt muons,
a $\lz$ with two tracks of the same type, ``long'' or ``downstream'',
built a common vertex and applied a constrained fit with the masses of the
$\psi$ and the $\lz$.

\begin{figure}[h]
\centering
\includegraphics[width=40mm]{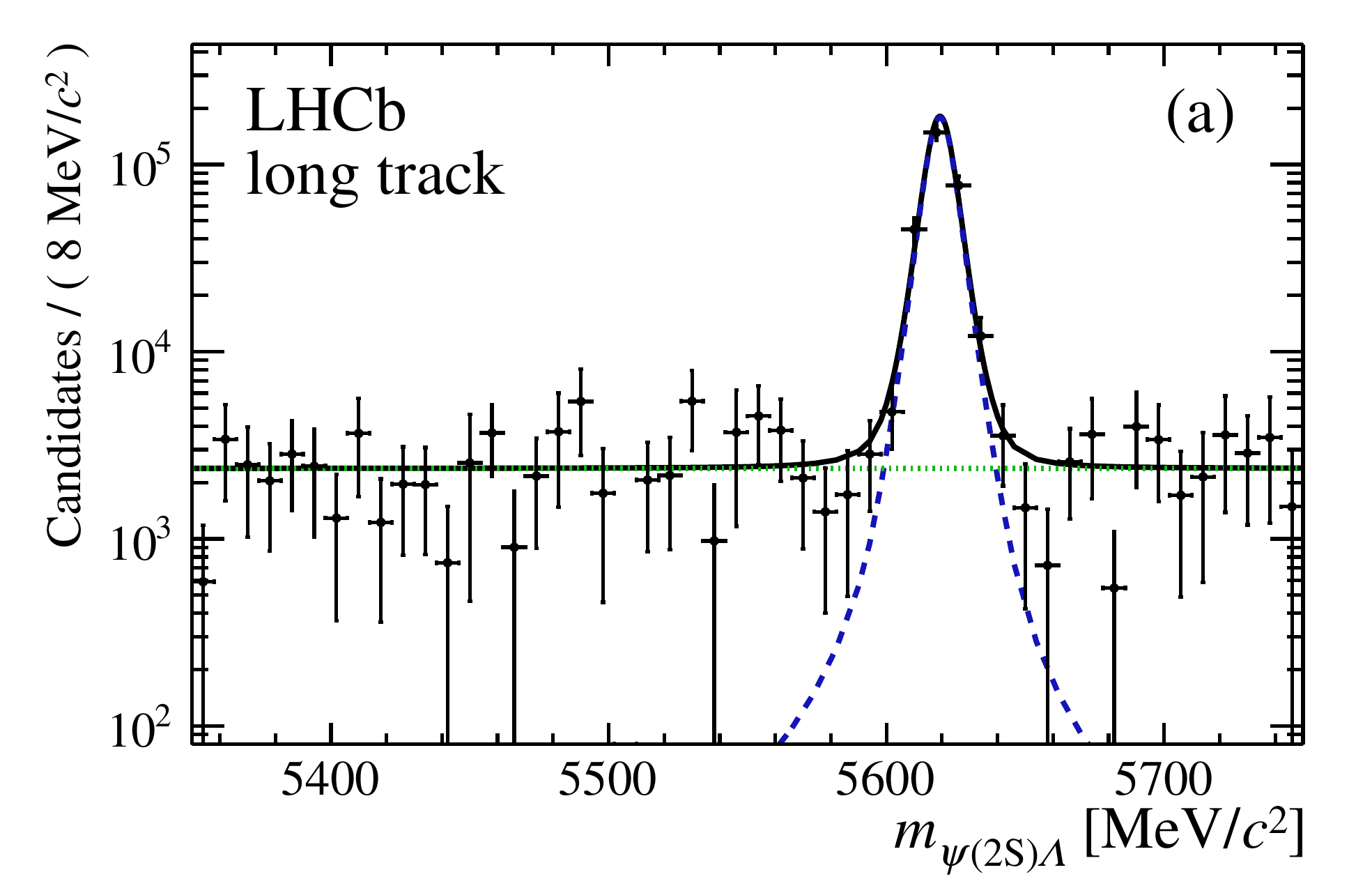}
\includegraphics[width=40mm]{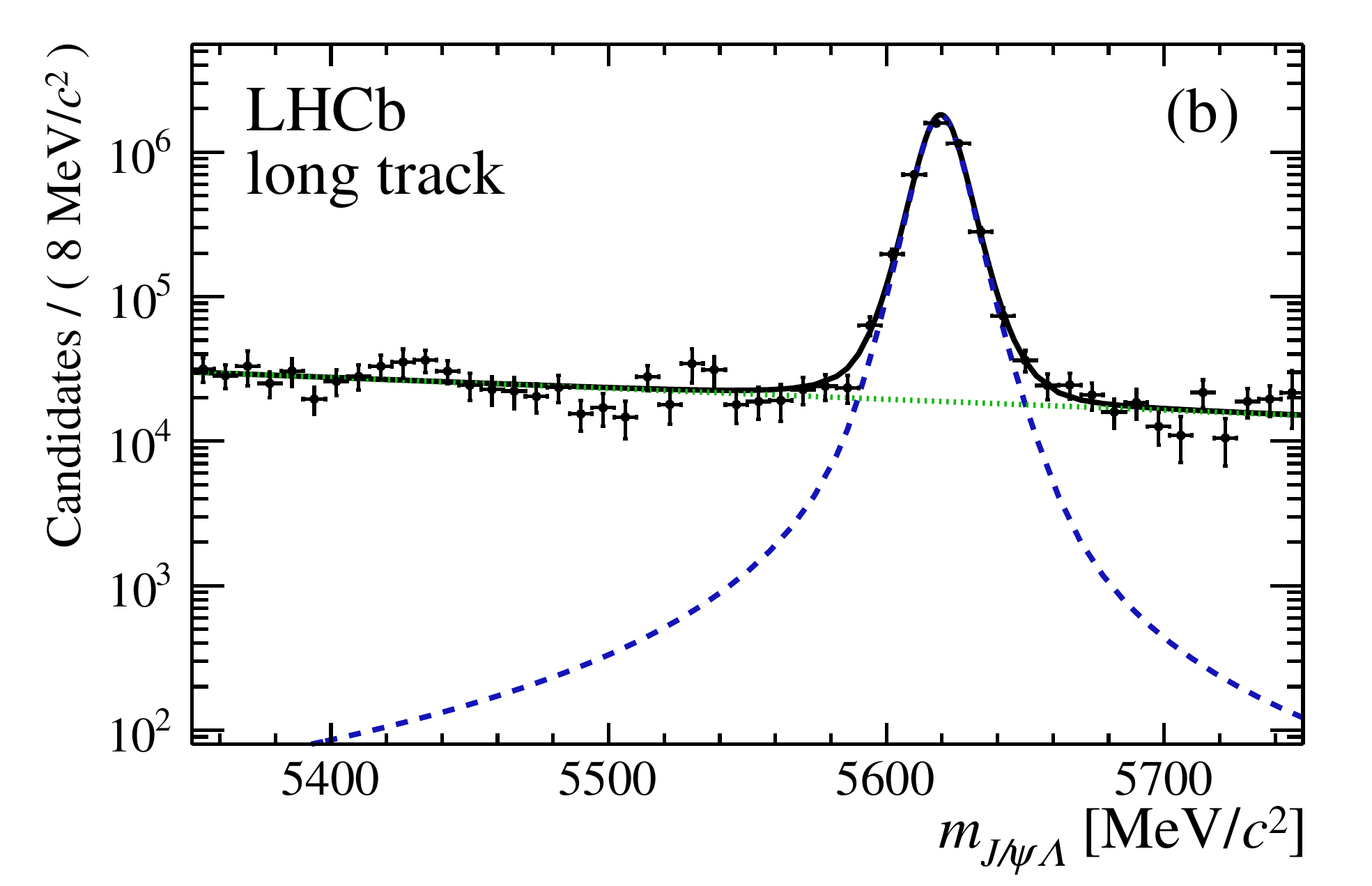}
\includegraphics[width=40mm]{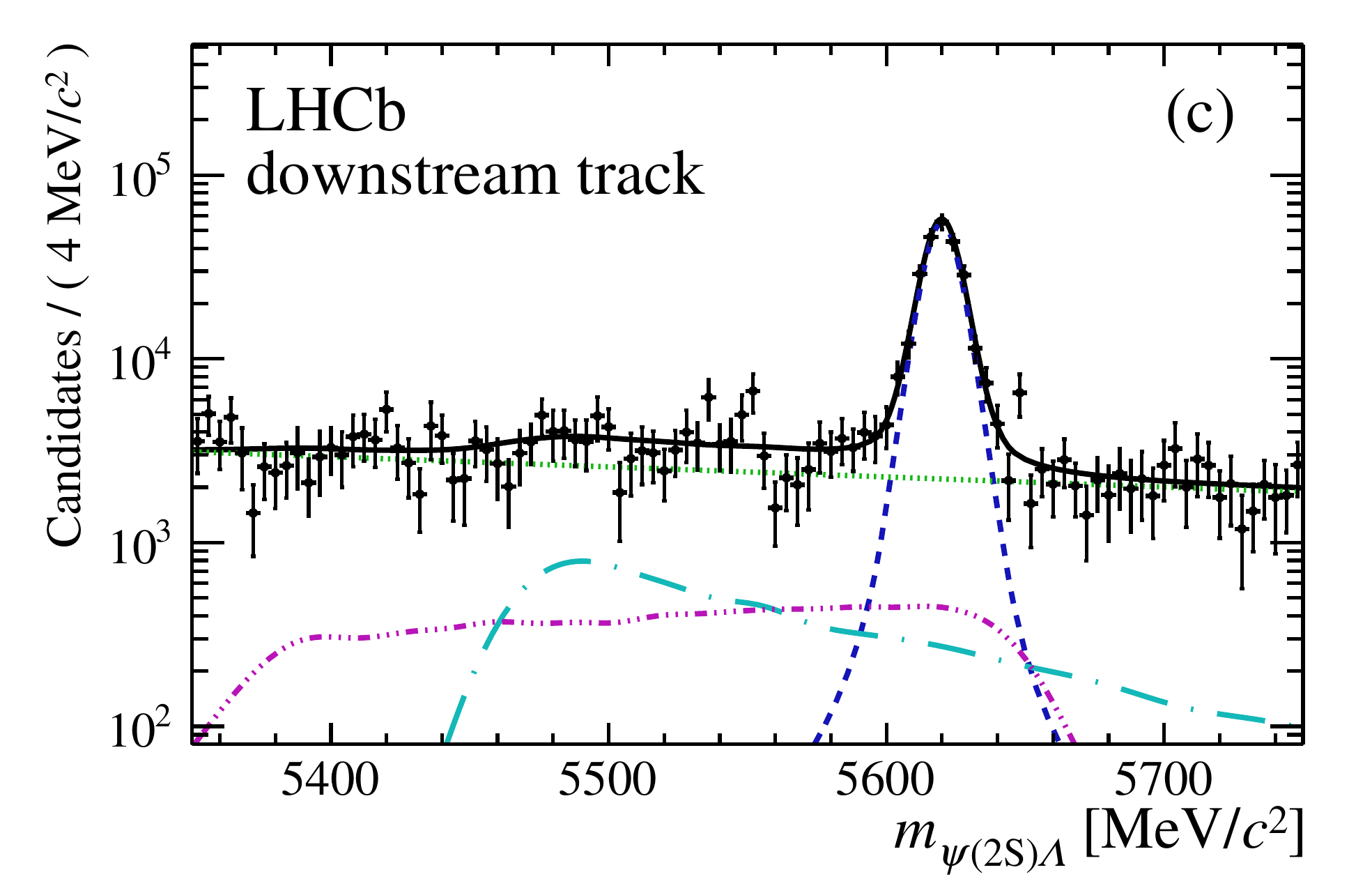}
\includegraphics[width=40mm]{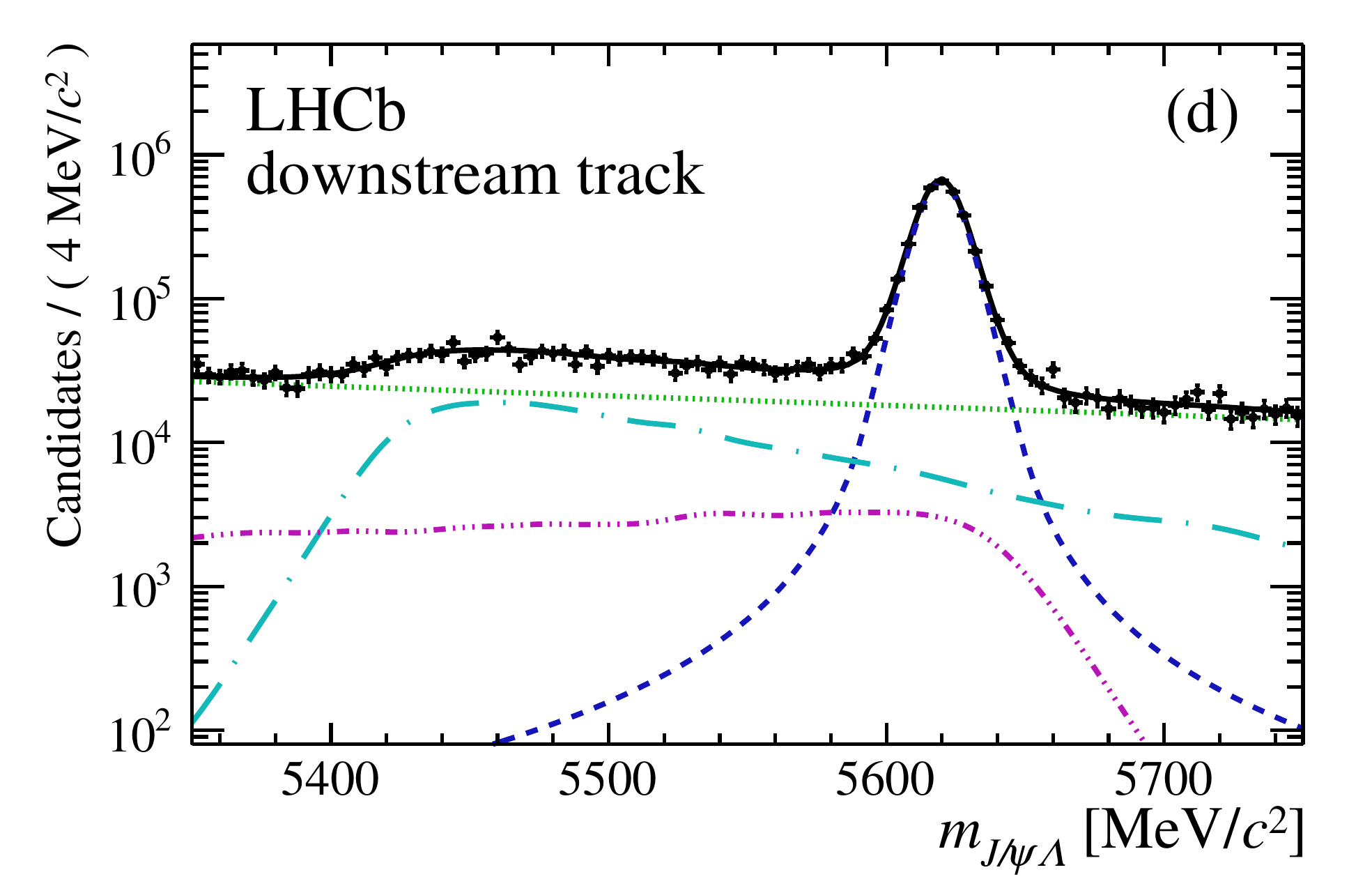}
\caption{Fits to the invariant-mass distributions for
  $\lb \rightarrow \jps{2} \lz$ (a,c) and $\lb \rightarrow \jpsi \lz$ (b,d)
  obtained by LHCb~\cite{ref:lBRLbr}. The signal (blue, dashed), the
  combinatorial background (green, dotted),
  the $\bz \rightarrow \psi \ks$ background (cyan, long-dash-dotted) and
  the $\xsbn \rightarrow \psi \Xi^-$ background (violet, dash-triple-dotted)
  are indicated.}
\label{fig:lBRLbr}
\end{figure}

%Events have been weighted with the inverse of efficiency that was
%estimated in the simulation as well as the background from decays of
%$\bz \rightarrow \psi (\ks \rightarrow \pi^+ \pi^-)$ or
%$\xsbn \rightarrow \psi (\Xi^- \rightarrow \lz \pi^-)$; the invariant
%mass distributions are shown in Fig.~\ref{fig:lBRLbr}.
Events have been weighted with the inverse of efficiency; the latter was
estimated in the simulation as well as the background from decays of
$\bz \rightarrow \psi (\ks \rightarrow \pi^+ \pi^-)$ or
$\xsbn \rightarrow \psi (\Xi^- \rightarrow \lz \pi^-)$: the invariant
mass distributions are shown in Fig.~\ref{fig:lBRLbr}.

Taking the signal yields from the mass distributions fit and 
branching fractions of the $\psi$ from PDG~\cite{ref:pdgPRD} the
branching fractions ratio was obtained:\newpage

\begin{eqnarray*}
  & & \frac{\Br(\lb \rightarrow \jps{2} \lz)}
           {\Br(\lb \rightarrow \jpsi \lz)} =\\
  & & 0.513 \pm 0.023 (\mathtx{stat}) \pm 0.016 (\mathtx{syst}) \pm 0.011 (\mathtx{b.r.})
\end{eqnarray*}

where the last
uncertainty comes from the $\psi$ branching fraction.

\subsection{Baryon production in meson decays}

\subsubsection{$\bq$ decay}

Some special interest related to baryons can be found in heavy hadron
decays not only when baryon themselves are decaying, but also when they are 
present in the final state. Their presence can be used to look for
possible pentaquark intermediate states; an evidence was claimed by
LHCb~\cite{ref:lLBpk1,ref:lLBpk2}
in the decay of $\lb \rightarrow \jpsi p K^-$.
The presence of a baryon and an antibaryon can also
test possible glueball states~\cite{ref:gball1,ref:gball2}.
LHCb then studied the decays
$\bq \rightarrow \jpsi p \bar{p}$~\cite{ref:lBJPpp}; both the decays are
suppressed: the $\bd$ decay in this channel is suppressed by Cabibbo while
the $\bs$ decay in the same channel is suppressed by OZI.
A branching fraction at the level of $10^{-9}$ would be expected, with some
enhancement via a resonant contribution from
$f_J(2220) \rightarrow p \bar{p}$.

%as shown in Fig.~{fig:bzJPpp}.
%
%\begin{figure}[h]
%\centering
%\includegraphics[width=60mm]{dqprl_.pdf}
%\caption{.}
%\label{fig:aDJPsi}
%\end{figure}

The branching fraction is measured by a comparison with a normalization
channel, so that the ratio of branching fractions is measured, using the
well known $\bsjphi$ decay as reference.
The branching fraction of the studied channel is given by the ratio
of reconstructed decays, multiplied by the branching fractions of
$\bsjphi$, $\phi \rightarrow \kk$ and, only for $\bd$,
the ratio of fragmentation fractions:

\begin{eqnarray*}
  & &\Br(\bd \rightarrow \jpsi \pp) =\\
  & &\frac{N_{\bd \rightarrow \jpsi \pp}}{N_{\bs \rightarrow \jpsi \phi}}
  \cdot \Br(\bsjphi) \cdot \Br(\phi \rightarrow \kk) \cdot \frac{f_s}{f_d}\\
  & &\Br(\bs \rightarrow \jpsi \pp) =\\
  & &\frac{N_{\bs \rightarrow \jpsi \pp}}{N_{\bs \rightarrow \jpsi \phi}}
  \cdot \Br(\bsjphi) \cdot \Br(\phi \rightarrow \kk)\makebox[0pt][l]{~.}
\end{eqnarray*}

\begin{figure}[h]
\centering
\includegraphics[width=60mm]{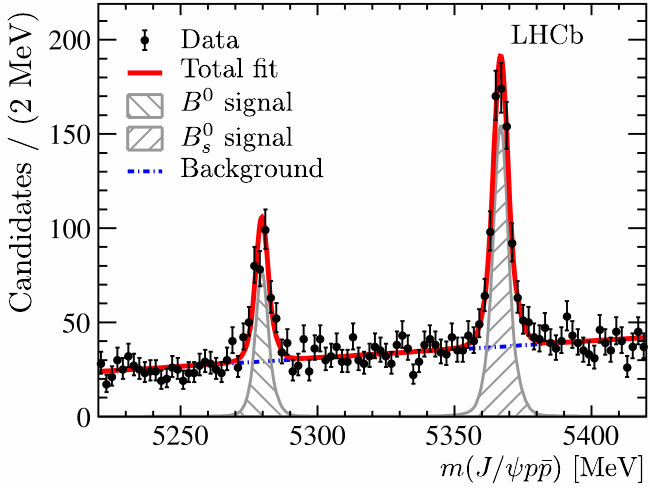}
\caption{Fit to invariant mass distribution of $\bdsjpp$ from
  LHCb~\cite{ref:lBJPpp}.}
\label{fig:lBJPpp}
\end{figure}

The number of events was obtained by an extended maximum likelihood fit
to the mass distributions, as shown in Fig.~\ref{fig:lBJPpp}; the product
$\Br(\bsjphi) \cdot \Br(\phi \rightarrow \kk) \cdot f_s / f_d$
was measured~\cite{ref:brProd} at $\eold$ as well as the fragmentation
ratio $f_s / f_d$~\cite{ref:lfsfdr,ref:lfsfdu}
and scaled to~$\elhc$~\cite{ref:lfsfds}.

The $\bdsjpp$ decay branching ratios have finally been extracted:\newpage

\begin{eqnarray*}
  & & \Br(\bd \rightarrow \jpsi \pp) =\\
  & & (4.51 \pm 0.40 (\mathtx{stat}) \pm 0.44 (\mathtx{syst})) \times 10^{-7}\\
  & & \Br(\bs \rightarrow \jpsi \pp) =\\
  & & (3.58 \pm 0.19 (\mathtx{stat}) \pm 0.33 (\mathtx{syst})) \times 10^{-6}\makebox[0pt][l]{~.}
\end{eqnarray*}

Due to the very low phase space available the momentum uncertainty is
negligible; that does allow as a side results the most precise single
measurements of $\bd$ and $\bs$ masses:

\begin{eqnarray*}
%  & & m_\bd =\\
%  & & (5279.74 \pm 0.30 (\mathtx{stat}) \pm 0.10 (\mathtx{syst}))~\mev \\
%  & & m_\bs =\\
%  & & (5366.85 \pm 0.19 (\mathtx{stat}) \pm 0.13 (\mathtx{syst}))~\mev
  & & m_\bd = (5279.74 \pm 0.30 (\mathtx{stat}) \pm 0.10 (\mathtx{syst}))~\mev\\
  & & m_\bs = (5366.85 \pm 0.19 (\mathtx{stat}) \pm 0.13 (\mathtx{syst}))~\mev\makebox[0pt][l]{~.}
\end{eqnarray*}

\subsubsection{$\bp$ decay}

Another study including the look for intermediate states has been done
by CMS about the decay $\bpjlbp$~\cite{ref:cBpJLp}: that decay was
first seen at $B$-factories~\cite{ref:bBaBar,ref:bBelle};
in this new study new exotic states were searched in the
$\jpsi \bar{\Lambda}^0$ or $\jpsi p$ systems.

As in the previous study of $\bq$ decay branching fraction has been
measured as ratio with the normalization channel
$\bp \rightarrow \jpsi K^{*+}$
($K^{*+} \rightarrow \ks \pi^+$~, $\ks \rightarrow \pi^+ \pi^-$).

The ratio and the absolute vaule of branching fractions were measured as:
\begin{eqnarray*}
  & & \frac{\brbjlp}
           {\Br(\bp \rightarrow \jpsi K^{*+})} =\\
  & & 1.054 \pm 0.057 (\mathtx{stat}) \pm 0.028 (\mathtx{syst}) \pm 0.011 (\mathtx{b.r.})\\
  & & \brbjlp =\\
  & & (15.07 \pm 0.81 (\mathtx{stat}) \pm 0.40 (\mathtx{syst}) \pm 0.86 (\mathtx{b.r.})) \times 10^{-6}
\end{eqnarray*}

where the last uncertainty comes from the involved cascade decays
branching fractions.

The distributions of invariant masses of the $\jpsi \bar{\Lambda}^0$
and $\jpsi p$ systems have then been studied and compared with expectations,
from pure phase space or phase space corrected for reflections from
$K^{*+} \rightarrow \bar{\Lambda}^0 p$ resonances. To do that the event
sample has been divided in $M(\bar{\Lambda}^0 p)$ invariant mass bins
and in each bin the first
8~Legendre polynomials and momenta have been computed using the
$\bar{\Lambda}^0 p$ helicity angle to describe the angular distributon.
Simulated events have then be reweighted using the $\bar{\Lambda}^0 p$
mass distribution ratio as reference, or the weights given by Legendre
polynomials and moments. The distributions obtained in this way have been
fitted to data.

\begin{figure}[h]
\centering
\includegraphics[width=60mm]{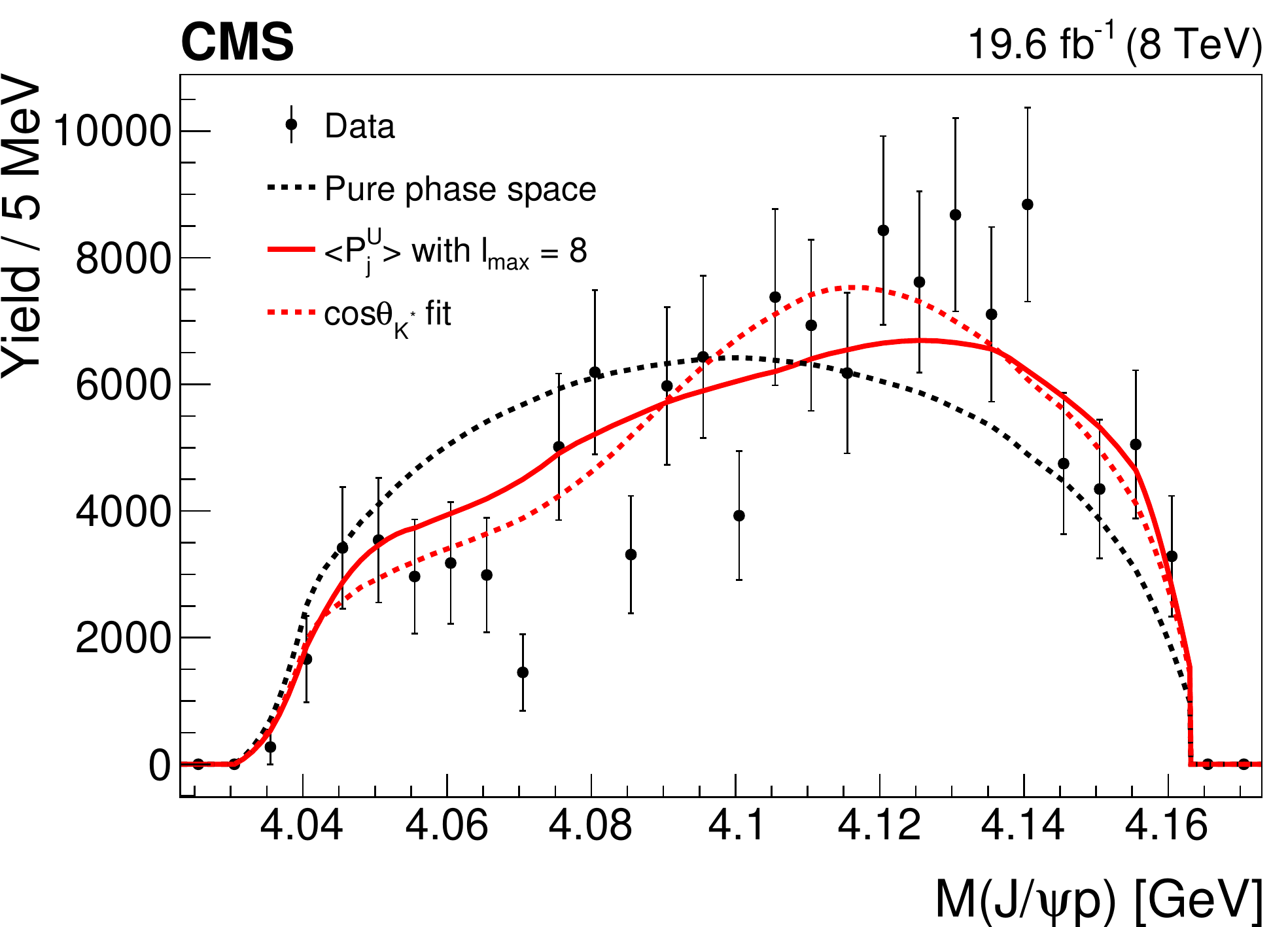}
\includegraphics[width=60mm]{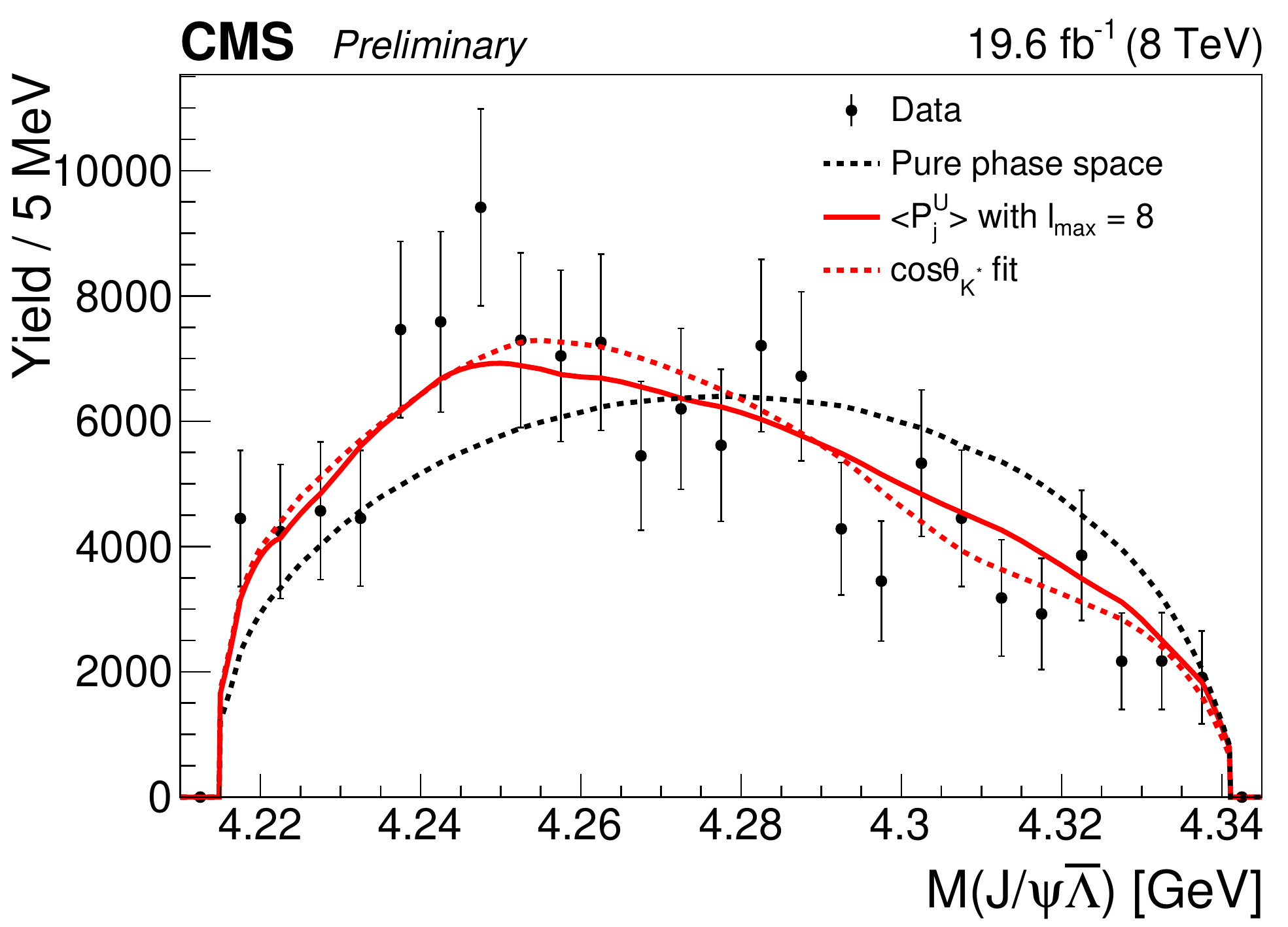}
\caption{Invariant mass distributions of $\jpsi p$ and
  $\jpsi \bar{\Lambda}^0$ obtaind by CMS~\cite{ref:cBpJLp} in the decay
  $\bpjlbp$, compared to the simulation using pure phase space (black),
  phase space corrected by the Legendre polynomials (red, solid)
  and a fit to the $\cos \theta_{K^*}$ distribution (red, dashed).}
\label{fig:cBpJLp}
\end{figure}

The $\jpsi p$ and $\jpsi \bar{\Lambda}^0$
invariant mass distributions are shown in Fig.~\ref{fig:cBpJLp},
compared with the simulation using pure phase space, the simulation
reweighted with the Legendre polynomials or a function fitted to the
$\cos \theta_{K^*}$ distribution in data.
%A large number of pseudoexperimentshave been generated according to
%the PDF for the pure phase space or the reweighted angular distributions;
%a log-likelihood ratio has then been computed, to extract a compatibility,
%or incompatibility, significance.

The quality of the data description from the different hypotheses has
been estimated generating a large number of pseudoexperiments according to
the PDF for the pure phase space or the reweighted angular distributions;
a log-likelihood ratio has then been computed, to extract a compatibility,
or incompatibility, significance. The significance of the incompatibility
of data with the pure phase space was found to be much larger than the
incompatibility with the phase space corrected by the Legendre polynomials:

\begin{center}
  \begin{tabular}{l|c|c|c}
    &
    $\jpsi p$ &
    $\jpsi \bar{\Lambda}^0$ &
    $\bar{\Lambda}^0 p$ \\ \hline
    pure phase-space &
    $5.5 \div 7.4$ &
    $6.1 \div 8.0$ &
    $3.4 \div 4.6$ \\
    reweighted phase-space &
    $1.3 \div 2.8$ &
    $1.3 \div 2.7$ &
    $ - $ 
  \end{tabular}
\end{center}

\section{Conclusions}
ATLAS, CMS and LHCb have produced many measurements of heavy hadron
production cross-sections and decay branching fractions:
\begin{itemize}
\item cross-sections have been compared to predictions and simulations
  and are input for other measurements,
\item branching fractions allow test model predictions,
\item in the study of decays the possible presence of
  intermediate exotic states has been investigated.
\end{itemize}
All those measurements allow important tests of QCD.

%% If you have acknowledgments, this puts in the proper section head.
%%\bigskip % extra skip inserted
%\begin{acknowledgments}
%This document is adapted from the ``Instruction for producing FPCP2003
%proceedings'' by P.~Perret 
%and from eConf templates~\cite{templates-ref}.
%\end{acknowledgments}

%\newcommand{\journalLink}{{\typeaut \aut}, {\typetit \tpr} \href{http://dx.doi.org/\DOI}{\typeref \jna {\typevol\vol}, (\yea) \pgn}}
%\newcommand{\journalDOI}{{\typeaut \aut}, {\typetit \tpr} {\typeref \jna {\typevol\vol} (\yea) \pgn}, {doi: \typedoi \dpr{\DOI}}}
%\newcommand{\journalNoLink}{{\typeaut \aut}, {\typetit \tpr} {\typeref \jna {\typevol\vol} (\yea) \pgn}}
%\newcommand{\prepriLink}{{\typeaut \aut}, {\typetit \tpr}\href{\URL} {\typeref \pgn}~(\yea)}
%\newcommand{\prepriDOI}{{\typeaut \aut}, {\typetit \tpr}\href{\URL} {\typeref \pgn~(\yea)}}
%\newcommand{\prepriNoLink}{{\typeaut \aut}, {\typetit \tpr} {\typeref \pgn~(\yea)}}
\newcommand{\journalLink}{{\typeaut\aut},{ }{\typetit\tpr}\href{http://dx.doi.org/\DOI}{\typeref\jna{ }{\typevol\vol},{ }\pgn{ }(\yea).}}
\newcommand{\journalDOI}{{\typeaut \aut},{ }{\typetit\tpr}{\typeref\jna{ }{\typevol\vol}{ }\pgn{ }(\yea)},{ }{doi: \typedoi\dpr{\DOI}}.}
\newcommand{\journalNoLink}{{\typeaut\aut},{ }{\typetit\tpr}{\typeref\jna{ }{\typevol\vol},{ }\pgn{ }(\yea)}.}
\newcommand{\prepriLink}{{\typeaut\aut},{ }{\typetit\tpr}\href{\URL}{ }{\typeref\pgn}.}
\newcommand{\prepriDOI}{{\typeaut\aut},{ }{\typetit\tpr}\href{\URL}{ }{\typeref\pgn}.}
\newcommand{\prepriNoLink}{{\typeaut\aut},{ }{\typetit\tpr}{\typeref\pgn}.}

\newcommand{\journal}{\journalNoLink}
\newcommand{\prepri}{\prepriNoLink}

\bigskip % extra skip inserted
%\begin{thebibliography}{9}   % Use for  1-9  references

\end{document}